\documentclass[epj,final]{svjour}
\usepackage{epsfig}
\begin{document}
\title{New experimental limits on the Pauli forbidden transitions in $^{12}$C nuclei obtained with 485 days Borexino data}
\author{G.~Bellini$^1$, J.~Benziger$^2$, S.~Bonetti$^1$, M.~Buizza Avanzini$^1$, B.~Caccianiga$^1$, L.~Cadonati$^3$, F.~Calaprice$^4$, C.~Carraro$^5$,
A.~Chavarria$^4$, F.~Dalnoki-Veress$^4$, D.~D'Angelo$^1$, S.~Davini$^5$, H.~de~Kerret$^6$, A.~Derbin$^7$, A.~Etenko$^8$,
K.~Fomenko$^9$, D.~Franco$^1$, C.~Galbiati$^4$, S.~Gazzana$^{10}$, C.~Ghiano$^{10}$, M.~Giammarchi$^1$, M.~Goeger-Neff$^{11}$,
A.~Goretti$^4$,, C.~Grieb$^{12}$, E.~Guardincerri$^5$, S.~Hardy$^{12}$, Aldo Ianni$^{10}$, Andrea Ianni$^4$, M.~Joyce$^{12}$,
G.~Korga$^{10}$, D.~Kryn$^6$, M.~Leung$^4$, T.~Lewke$^{11}$, E.~Litvinovich$^8$, B.~Loer$^4$, P.~Lombardi$^1$, L.~Ludhova$^1$,
I.~Machulin$^8$, S.~Manecki$^{12}$, W.~Maneschg$^{13}$, G.~Manuzio$^5$, Q.~Meindl$^{11}$, E.~Meroni$^1$, L.~Miramonti$^1$,
M.~Misiaszek$^{15,10}$, D.~Montanari$^{10,4}$, V.~Muratova$^7$, L.~Oberauer$^{11}$, M.~Obolensky$^6$, F.~Ortica$^{14}$,
M.~Pallavicini$^5$, L.~Papp$^{12}$, L.~Perasso$^1$, S.~Perasso$^5$, A.~Pocar$^3$, R.S.~Raghavan$^{12}$, G.~Ranucci$^1$,
A.~Razeto$^{10}$, A.~Re$^1$, P.~Risso$^5$, A.~Romani$^{14}$, D.~Rountree$^{12}$, A.~Sabelnikov$^8$, R.~Saldanha$^4$,
C.~Salvo$^5$, S.~Schoenert$^{13}$, H.~Simgen$^{13}$, M.~Skorokhvatov$^8$, O.~Smirnov$^9$, A.~Sotnikov$^9$, S.~Sukhotin$^8$,
Y.~Suvorov$^{1,8}$, R.~Tartaglia$^{10}$, G.~Testera$^5$, D.~Vignaud$^6$, R.B.~Vogelaar$^{12}$, F.~von Feilitzsch$^{11}$,
Y.~Winter$^{11}$, M.~Wojcik$^{15}$, M.~Wurm$^{11}$, J.~Xu$^4$, O.~Zaimidoroga$^9$, S.~Zavatarelli$^5$, G.~Zuzel$^{13}$ }

\institute{$^1$ Dipartimento di Fisica, Universita' degli Studi e INFN, 20133 Milano, Italy
\\ $^2$ Chemical Engineering Department, Princeton University, Princeton, NJ 08544, USA
\\ $^3$ Physics Department, University of Massachusetts, Amherst, AM01003, USA
\\ $^4$ Physics Department, Princeton University, Princeton, NJ 08544, USA
\\ $^5$ Dipartimento di Fisica, Universita' e INFN, Genova 16146, Italy
\\ $^6$ Laboratoire AstroParticule et Cosmologie, 75231 Paris cedex 13, France
\\ $^7$ St. Petersburg Nuclear Physics Institute, 188350 Gatchina, Russia
\\ $^8$ RRC Kurchatov Institute, 123182 Moscow, Russia
\\ $^9$ Joint Institute for Nuclear Research, 141980 Dubna, Russia
\\ $^{10}$ INFN Laboratori Nazionali del Gran Sasso, SS 17 bis Km 18+910, 67010 Assergi (AQ), Italy
\\ $^{11}$  Physik Department, Technische Universitaet Muenchen, 85747 Garching, Germany
\\ $^{12}$ Physics Department, Virginia Polytechnic Institute and State University, Blacksburg, VA 24061, USA
\\ $^{13}$ Max-Plank-Institut fuer Kernphysik, 69029 Heidelberg, Germany
\\ $^{14}$ Dipartimento di Chimica, Universita' e INFN, 06123 Perugia, Italy
\\ $^{15}$ M. Smoluchowski Institute of Physics, Jagellonian University, 30059 Krakow, Poland
\\
\\ Borexino collaboration}
\abstract{{ The Pauli exclusion principle (PEP) has been tested for nucleons ($n,p$) in $^{12}C$ with the Borexino detector.The
approach consists of a search for $\gamma$, $n$, $p$ and $\beta^\pm$ emitted in a non-Paulian transition of 1$P_{3/2}$- shell
nucleons to the filled 1$S_{1/2}$ shell in nuclei. Due to the extremely low background and the large mass (278 t) of the
Borexino detector, the following most stringent up-to-date experimental bounds on PEP violating transitions of nucleons have
been established: $\tau({^{12}\rm{C}}\rightarrow{^{12}\widetilde{\rm{C}}}+\gamma) \geq 5.0\cdot10^{31}$ y,
$\tau({^{12}\rm{C}}\rightarrow{^{11}\widetilde{\rm{B}}}+ p) \geq 8.9\cdot10^{29}$ y,
$\tau({^{12}\rm{C}}\rightarrow{^{11}\widetilde{\rm{C}}}+ n) \geq 3.4 \cdot 10^{30}$ y,
$\tau({^{12}\rm{C}}\rightarrow{^{12}\widetilde{\rm{N}}}+ e^- + \widetilde{\nu_e}) \geq 3.1\cdot 10^{30}$ y and
$\tau({^{12}\rm{C}}\rightarrow{^{12}\widetilde{\rm{B}}}+ e^+ + \nu_e) \geq 2.1 \cdot 10^{30}$ y, all at 90\% C.L. The
corresponding upper limits on the relative strengths for the searched non-Paulian electromagnetic, strong and weak transitions
have been estimated: $\delta^2_{\gamma} \leq 2.2\cdot10^{-57}$, $\delta^2_{N} \leq 4.1\cdot10^{-60 }$ and $\delta^2_{\beta} \leq
2.1\cdot10^{-35}$.} }

\titlerunning{New experimental limits on the Pauli....}
\authorrunning{Borexino coll.,}
\maketitle

\section{Introduction}
%%%%%%%%%%%%%%%%%%%%%%%%%%%%%%%%%%%%%%%%%%%%%%%%%%%%%%%%%%%%%%%%%%%%%%%%%%%%%%%%%%%%%%%%%%%%%%%

The exclusion principle was formulated by W.Pauli in 1925 and in its original form postulated that " these can never be two or
more equivalent electrons in an atom" \cite{Pau25}.  In the case of Bohr atoms it meant that only one electron with definite spin
orientation can occupy each of the allowed orbits.  This statement was later formalized in the framework of quantum-mechanics by
saying that for two identical electrons the total wave function is anti-symmetric under electron permutation. In relativistic
Quantum Field Theory (QFT), the Pauli Exclusion Principle (PEP) appears automatically for systems of identical fermions as a result
of the anti-commutativity of the fermion creation and annihilation operators.

Although PEP is of fundamental importance, its physical cause is not yet understood. According to Okun  "a non-conformist approach
to the PEP could be traced to Dirac and Fermi" \cite{Oku89}. Both Dirac and Fermi discussed the implications of a small PEP violation
on atomic transitions and on atomic properties \cite{Dir58,Fer33}.

The experimental searches for the possible PEP violations started about 15 years later when the electron stability was tested. Pioneering
experiments were performed by Reines and Sobel by searching for X-rays emitted in the transition of an L-shell electron to the filled
K-shell in an atom \cite{Rei74}, and by Logan and Ljubicic, who searched for $\gamma$-quanta emitted in a PEP-forbidden transition of
nucleons in nuclei \cite{Log79}.

In 1987-91 theoretical models  implicating  PEP violation were constructed by Ignatiev and Kuzmin \cite{Ign87}, Greenberg and
Mohapatra~\cite{Gre87,Gre90,Moh90} and by Okun \cite{Oku87}, but it was shown by Govorkov  \cite{Gov89} that even a small PEP-violation leads to
negative probabilities for some processes. Moreover, in 1980 Amado and Primakoff pointed out that in the framework of quantum mechanics
PEP-violating transitions \cite{Rei74,Log79} are forbidden even if PEP-violation takes place \cite{Ama80}.

At present, no acceptable theoretical formalism exists. In particular, it is not possible to account for PEP violation by means of a
self-consistent and non-contradictory "small" parameter, as in the case of $P$- and $CP$-symmetry violation or $L$- and $B$-
non-conservation. The results of experiments are presented as lifetime limits or as limits on the relative strength of the normal and
Pauli-forbidden transitions. Critical studies of the possible violation of PEP have been done both theoretically and experimentally in
\cite{Oku89,Oku89A,Ign05}. More reviews and references can be found in \cite{Hil00,SSC08}.

There are two (or four, if we consider electrons and nucleons separately) types of experiments to look for PEP violation. The
first one is based on the search for atoms or nuclei in a non-Paulian state; the second one is based on the search for the
prompt radiation accompanying non-Paulian transitions of electrons or nucleons.

Experiments of the first type have been performed by Novikov et al. \cite{Nov89,Nov90} and Nolte et al. \cite{Nol91} who looked for
non-Paulian exotic atoms of $^{20}$Ne and $^{36}$Ar with 3 electrons on K-shell using mass spectroscopy on fluorine and chlorine samples.
Similarly, the atoms of Be with 4 electrons in 1s-state that look like He atoms were searched for by Javorsek et al. \cite{Jav00}. The
anomalous carbon atoms in boron samples were searched for by $\gamma$-activation analysis in \cite{Bar98} (Barabash et al.). The
PEP-forbidden nuclei of $^{5}$Li with 3 protons in the 1S-shell was searched for by Nolte et al., using time-of-flight mass spectroscopy~\cite{Nol90}.

Goldhaber was the first to point out that the same experimental data which were used to set a limit on the lifetime of the
electron can be used to test the validity of the PEP for atomic electrons \cite{Rei74}. From the experimental point of view, the
searches for characteristic X-rays due to electron decay inside an atomic shell \cite{Fei59}-\cite{DAMA09} are often
indistinguishable from the PEP-violating transition, but according to Amado and Primakoff {\cite{Ama80} these transition do not
take place even if PEP is violated. This restriction is not valid for transitions accompanied by a change of the number of
identical fermions (e.g. non-Paulian $\beta^{\pm}$-transitions) and can be evaded in composite models of electron or models
including extra dimensions \cite{Gre87,Aka92}.

The new method was realized by Ramberg and Snow which looked for anomalous X-rays emitted by Cu atoms in a conductor \cite{Ram88}. The
established upper limit on the probability for the 'new' electron passing in the conductor to form a non-Paulian atom with 3 electrons
in the K-shell is $1.7\cdot10^{-26}$. An improvement of the sensitivity of the method is currently being planned by the VIP collaboration
\cite{Bar06}.  Laser atomic and molecular spectroscopy were used to search for anomalous PEP- forbidden spectral lines of
$^4$He atoms \cite{Dei95} (Deilamian et al.) and molecules of O$_2$ \cite{Hil96,Ang96} (Hilborn et al., Angelis et al.,) and CO$_2$
\cite{Mod98} (Modugno et al.).

The violation of PEP in the nucleon system has been studied by searching for the non-Paulian transitions with $\gamma$- emission
\cite{KAMIOKANDE,NEMO} (Kamiokande, NEMO-II), $p$-emission \cite{Ejiri,DAMA,DAMA09} (Elegant-V, DAMA/LIBRA) and $n$-emission
\cite{Kishimoto} (Koshomoto et al.,), non-Paulian $\beta^+$- and $\beta^-$- decays \cite{Kek90,NEMO}, (LSD, Kekez et al., NEMO-II) and in
nuclear $(p,p)$, $(p,\alpha)$- reactions on $^{12}$C \cite{pp-reaction} (Miljani et al.).

The strongest limits for non-Paulian transitions in $^{12}$C with $\gamma$-,~$p$-,~$n$-,~$\alpha$-, and $\beta^{\pm}$- emissions
were obtained with a prototype of the Borexino detector - Counting Test Facility (CTF)~\cite{Bor04}.
%Although the majority of the lifetime limits obtained with CTF are still the best up-to-date, the PEP is one of the most fundamental laws of nature and it has to be valid to the extent determined by experiment.
In this letter we present the new results obtained with 485 days of Borexino data. The large Borexino mass (70 times larger than
the CTF one) and its extremely low background level (200 times lower than in CTF at 2 MeV) enabled us to improve the lifetime
limits for non-Paulian transitions in $^{12}\rm{C}$ by 3-4 orders of magnitude with respect to CTF.

\section{Experimental set-up and measurements}
%%%%%%%%%%%%%%%%%%%%%%%%%%%%%%%%%%%%%%%%%%%%%%%%%%%%%%%%%%%%%%%%%%%%%%%%%%%%%%%%%%%%%%%%%%%%%%%
\subsection {Brief description of Borexino}
%%-------------------------------------------------------------------------------------
Borexino is a real-time detector for solar neutrino spectroscopy located at the Gran Sasso Underground Laboratory. Its main goal
is to measure low energy solar neutrinos via ($\nu$,e)-scattering in an ultra-pure liquid scintillator. The extremely high radiopurity of
the detector and its large mass allow to simultaneously address other fundamental questions particle physics and astrophysics.

The main features of the Borexino detector and its components have been thoroughly described in \cite{Ali02}-\cite{Ali08}.
Borexino is a scintillator detector with an active mass of 278 tons of pseudocumene (PC, C$_9$H$_{12}$), doped with 1.5 g/liter
of PPO (C$_{15}$H$_{11}$NO). The scintillator is inside a thin nylon vessel (IV - inner vessel) and is surrounded by two
concentric PC buffers (323 and 567 tons) doped with a small amount of light quencher (dymethilpthalate-DMP) to reduce their
scintillation. The two PC buffers are separated by a second thin nylon membrane to prevent diffusion of radon coming from PMTs,
light concentrators and SSS walls towards the scintillator. The scintillator and buffers are contained in a Stainless Steel
Sphere (SSS) with diameter 13.7 m. The SSS is enclosed in a 18.0-m diameter, 16.9-m high domed Water Tank (WT), containing 2100
tons of ultrapure water as an additional shield against external $\gamma$'s and neutrons. The scintillation light is detected by
2212 8" PMTs uniformly distributed on the inner surface of the SSS. All the internal components of the detector were selected
following stringent radiopurity criteria. The WT is equipped with 208 additional PMTs that act as a Cerenkov muon detector
(outer detector) to identify the residual muons crossing the detector.

\subsection{Detector calibration. Energy and spatial resolutions.}
%%----------------------------------------------------------------------------------------------
In Borexino charged particles are detected by their scintillation light-producing interactions with the liquid scintillator. The energy
of an event is measured using the total collected light from all PMT's. In a simple approach, the
response of the detector is assumed to be linear with respect to the energy released in the scintillator. The coefficient
linking the event energy and the total collected charge is called the light yield (or photoelectron yield). Deviations from
linearity at low energy can be taken into account by the ionization deficit function $ f(k_{B},E) $, where $k_B$ is the empirical Birks' constant \cite{Bir51}.

\begin{figure}
\includegraphics[bb = 30 90 500 760, width=8cm,height=10cm]{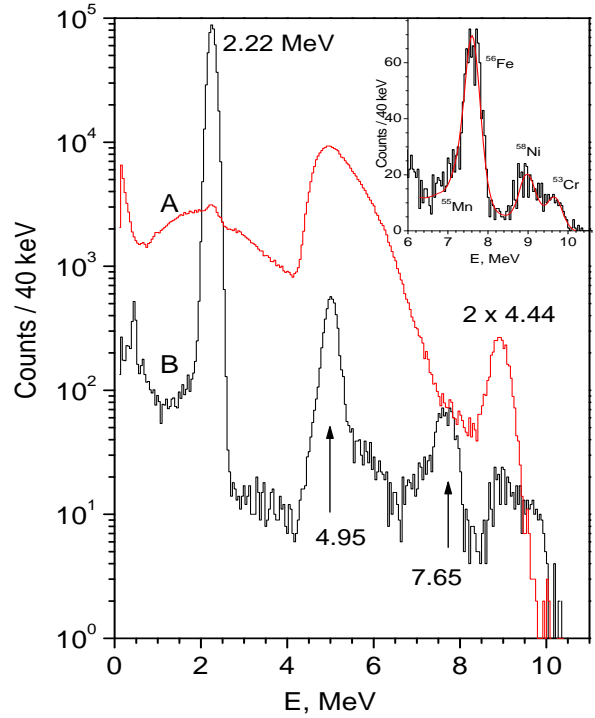}
\caption { The energy spectra of prompt (A) and delayed (B) signals registered with $^{241}$Am$^9$Be source. In insert, the
$\gamma$-lines from neutron captures on stainless steel holder of AmBe-source are shown. } \label{AmBe}
\end{figure}

\begin{figure}
\includegraphics[bb = 30 90 500 760, width=8cm,height=10cm]{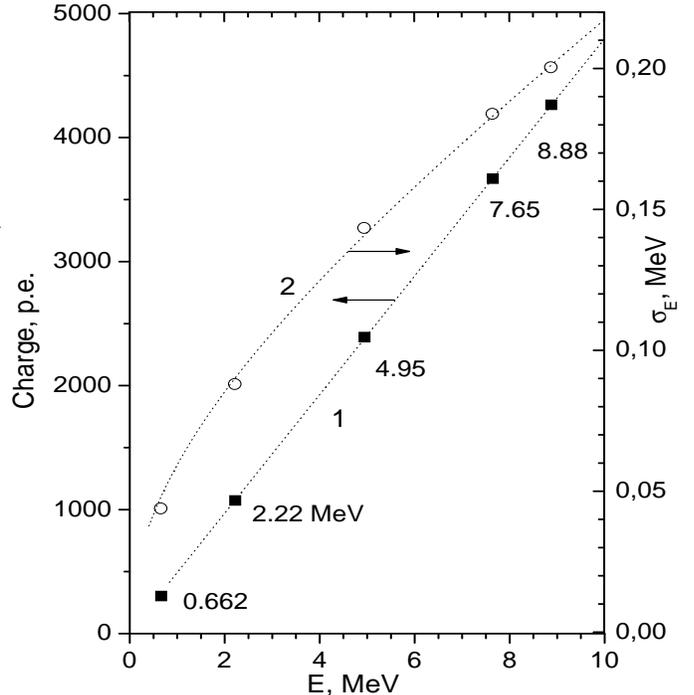}
\caption { The dependency of registered charge vs energy of $\gamma$-quanta (squares, left scale). The corresponding energy
resolution ($\sigma_E$) is indicated on the right scale (cycles). } \label{Calib}
\end{figure}

The detector energy and spatial resolution were studied with radioactive sources placed at different positions inside the inner
vessel. For relatively high  energies ( $>$2 MeV), which are of interest for non-Paulian transition studies, the energy
calibration was performed with a $^{241}$Am-$^9$Be neutron source. Fig.\ref{AmBe} shows the spectrum obtained with the source
placed at the center of the detector. The reactions ${^9\rm{Be}}(\alpha,n){^{12}\rm{C_{gs}}}$ and
$^9\rm{Be}(\alpha,\rm{n})^{12}\rm{C^*}$ (4.44 MeV) produce two main neutron groups with energies up to 11 MeV and 6.5 MeV,
respectively. The resulting neutrons are thermalized by elastic and inelastic scattering in the hydrogen-rich organic
scintillator and eventually are captured by protons or carbon nuclei. The upper (red) spectrum in Fig.\ref{AmBe} corresponds to
the prompt neutrons and $\gamma$'s, while the lower (black) one is that of the delayed signals. The energy scale was determined
with the 2.22 MeV and 4.95 MeV $\gamma$ de-excitations following neutron capture on $^1$H and $^{12}$C nuclei, and with the 8.88
MeV peak, sum of two 4.44 MeV $\gamma$ quanta. The expected shift of the 8.88 MeV peak position (due to residual energy of the
scattered neutron) is suppressed by the sizable quenching factor of low energy protons.  The 7.65 MeV $\gamma$-line following
neutron capture on $^{56}$Fe present in the source holder was used also. The deviations from linearity of the $\gamma$-peaks
positions was less than 30 keV over the whole range. The energy resolution scales approximately as ${(\sigma /E)}$ $\simeq
(0.058+1.1\cdot10^{-3}E)/\sqrt{E}$ where E is given in MeV (Fig.\ref{Calib}). The position of an event is determined using a
photon time of flight reconstruction algorithm. The resolution in the event position reconstruction is 13$\pm$2 cm in the x and
y coordinates, and 14$\pm$2 cm in z, measured with the $^{214}$Bi-$^{214}$Po $\beta-\alpha$ decay sequence.

\section{Data analysis}
%%%%%%%%%%%%%%%%%%%%%%%%%%%%%%%%%%%%%%%%%%%%%%%%%%%%%%%%%%%%%%%%%%%%%%%%%%%%%%%%%%%%%%%%%%%%%%%
\subsection{Theoretical considerations}
%%---------------------------------------------------------------------------------------------
The non-Paulian transitions were searched for in $^{12}$C  nuclei of the PC. The nucleon level scheme of $^{12}$C in
a simple shell model is shown in Fig.\ref{Schema}. The non-Paulian transitions which have been searched for in the analysis
described in this paper are schematically illustrated. The transition of a nucleon from the $P$-shell to the filled $S$-shell
will result in excited non-Paulian nuclei $^{12}\widetilde{C}$. The excitation energy corresponds to the difference of the
binding energies of nucleons on $S$- and $P$-shells and is comparable with the separation energies of protons $S_p$, neutrons
$S_n$ and $\alpha$-particles $S_{\alpha}$. Hence, together with the emission of $\gamma$-quanta, the emission of $n$, $p$ and
$\alpha$ is possible. In this paper we also discuss weak processes violating PEP, like ${\beta}^{+}$- and ${\beta}^{-}$-decay to
a non-Paulian nucleon in the final $1S_{1/2}$-state.

The energy released in the transitions under consideration is the difference between the binding energies of the final and
initial nuclei:
\begin{center}
$Q(^{12}\rm{C}\rightarrow \widetilde{X}+Y)=M(^{12}\rm{C})-M(\widetilde{X})-M(Y)=$
\end{center}
\begin{equation}
-E_{b}(^{12}\rm{C})+E_{b}(\widetilde{X})+E_{b}(Y);
\end{equation}
where $\widetilde{X}$ denotes a non-Paulian nucleus, Y = $\gamma, p, n, d, \alpha..$ is the particle or nucleus emitted and
$E_b$ is the corresponding binding energies which are well known for normal nuclei \cite{Aud95}. The signature of non-Paulian
transitions with two particles in the final state is a peak in the experimental spectrum with the width defined by the energy
resolution of the detector.

\begin{figure}
\includegraphics[bb = 30 90 500 760, width=8cm,height=8cm]{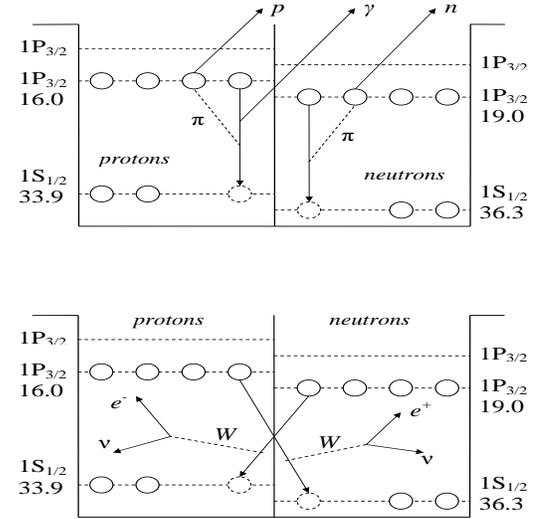}
\caption {Occupation of energy levels by protons and neutrons for the $^{12}$C ground state in a simple shell model. Schemes of
non-Paulian transitions of nucleons from the $P$-shell to the filled $S$-shell: a) with $\gamma$-, $n$-, $p$- and
$\alpha$-emission; b) with $\beta^+$-, $\beta^-$-emission.} \label{Schema}
\end{figure}

In the case of non-Paulian transitions induced by weak interactions, the $\beta^{\pm}$-spectra have to be observed. The end-point energy of
the $\beta$-spectrum in the reaction $ ^{12}$C$\rightarrow ^{12}\widetilde{N}+e^{-}+\overline{\nu} $ is
\begin{equation}
Q=m_{n}-m_{p}-m_{e}-E_{b}(^{12}\rm{C})+E_{b}(^{12}\widetilde{\rm{N}}).
\end{equation}
A similar equation can be written for non-Paulian transition with $\beta^+$-emission, but the registered energy will be shifted
by $\approx2m_e$ due to positron annihilation quanta.

The binding energy of the non-Paulian nuclei with 3 neutrons or 3 protons  on the $1S_{1/2}$-shell $E_{b}(\widetilde{X})$ can be evaluated
considering the binding energy of normal nuclei $ E_{b}(X) $ and the difference between the binding energies of nucleons on the
$1S_{1/2}$-shell $ E_{n,p}(S_{1/2}) $ and the binding energy of the last nucleon $ S_{n,p}(X) $:
\begin{equation}
E_{b}(\widetilde{X}_{n,p})\simeq E_{b}(X)+E_{n,p}(1S_{1/2})-S_{n,p}(X).
\end{equation}

The nucleon binding energies for light nuclei ($^{12}$C, $^{11}$B and others) were measured while studying $(p,2p)$ and $(p,np)$
proton scattering reactions with 1 GeV energy at PNPI proton synchrocyclotron \cite{PNPI}. Using these data we calculated the
Q-values (with errors) for different non-Paulian transitions which are shown in Table~\ref{Qvalues}. The details of the
calculations can be found in our previous work \cite{Bor04}.
%%%%%%%%%%%%%%%%%%%%%%%%%%%%%%%%%%%%%%%%%%%%%% TABLE 1 %%%%%%%%%%%%%%%%%%%%%%%%%%%%%%%%%%%%%%%%%%%%%%%%%
\begin{table}[!htbp]
\begin{center}
\caption{ The energies released in the transitions with non-Paulian nuclei with 3 neutrons or 3 protons  on the S-shell in the final
state.}\label{Qvalues}
\begin{tabular}{|l|c|c|}
  \hline
    Channel                                     & $Q_3p$,         & $Q_3n$       \\
                                                & (MeV)           & (MeV)        \\ \hline
   $^{12}\rm{C}\rightarrow{^{12}\widetilde{\rm{C}}}+\gamma$ & $17.9\pm0.9$    & $17.7\pm0.6$ \\

   $^{12}\rm{C}\rightarrow{^{11}\widetilde{\rm{B}}}+ p$     & $6.3\pm0.9 $    & $7.8\pm1.0 $ \\
   $^{12}\rm{C}\rightarrow{^{11}\widetilde{\rm{C}}}+ n$     & $6.5\pm0.9 $    & $4.5\pm0.6 $  \\
   $^{12}\rm{C}\rightarrow{^{8}\widetilde{\rm{Be}}}+\alpha$ & $3.0\pm0.6 $    &  $2.9\pm0.9  $  \\
   $^{12}\rm{C}\rightarrow{^{12}\widetilde{\rm{N}}}+ e^{-}+\overline{\nu_{e}}$ & $18.9\pm0.9 $ & - \\
   $^{12}\rm{C}\rightarrow{^{12}\widetilde{\rm{B}}}+ e^{+}+\nu_{e}$            & -             & $17.8\pm0.9 $  \\
   \hline
 \end{tabular}
\end{center}
\end{table}

For all other reactions such as $^{12}\rm{C}\rightarrow{^{10}\widetilde{\rm{B}}}+ d$,
$^{12}\rm{C}\rightarrow{^{9}\widetilde{\rm{B}}}+ t$, ${^{12}\rm{C}}\rightarrow{^{9}\widetilde{\rm{Be}}}+ {^3\rm{He}}$,
${^{12}\rm{C}}\rightarrow{^{6}\widetilde{\rm{Li}}}+ {^6\rm{Li}}$ and ${^{12}\rm{C}}\rightarrow{^{6}\widetilde{\rm{Li}}}+
{^4\rm{He}} + d$, except in the process ${^{12}\rm{C}}\rightarrow{^{9}\widetilde{\rm{B_{3p}}}}+ t$, the Q-values are negative.

Using the obtained Q-values one can calculate the detector response for all the reactions mentioned above. The recoil energy of
nuclei and quenching factors for different particles have to be taken into account.

Because of the uncertainties in the non-Paulian nuclei properties, the prediction of the branching ratio for the emission in
each of the above mentioned channels has a poor significance. For the case of the neutron disappearance (e.g. invisible decay
$n\rightarrow 3\nu$) from the $1S_{1/2}$-shell in $^{12}$C nuclei, the branching ratio and spectra of the emitted particles were
considered in \cite{Kam02}. For the excitation energy of $^{11}$C of 17 MeV they found that the branching ratios for $p$-, $n$-,
and $\alpha$-emission are of the same order of magnitude and, it is negligible for $\gamma$-emission. In the present paper we
give the separate limits on the probabilities for each of the non-Paulian reactions. Then, we compare the obtained results with
the corresponding rates of normal transitions.

\subsection{Data selection}
%%---------------------------------------------------------------------------------------------
Candidate events are selected by the following criteria: (1) events must have a unique cluster of PMT hits; (2) events should
not be flagged as muons by the outer Cherenkov detector; (3) events should not follow a muon within a time window of 2 ms;
(4)events should not be followed by another event within a time window of 2 ms except in case of neutron emission; (5) events
must be reconstructed within the detector volume. Depending on the specific channel under study, pulse-shape-discrimination has
also been applied to select events induced by $\gamma$, $\beta$, $p$ or $\alpha$.

The experimental energy spectra of Borexino in the range 1.0-14 MeV, collected during 485~days of data-taking (live time), is
shown in Fig.\ref{Spectr_all}.  The raw spectrum is presented at the top. At energies below 3~MeV, the spectrum is dominated by
2.6 MeV $\gamma$'s from the $\beta$-decay of $^{208}$Tl due to radioactive contaminations in the PMTs and in the SSS.

\begin{figure}
\includegraphics[bb = 30 90 500 760, width=8cm,height=10cm]{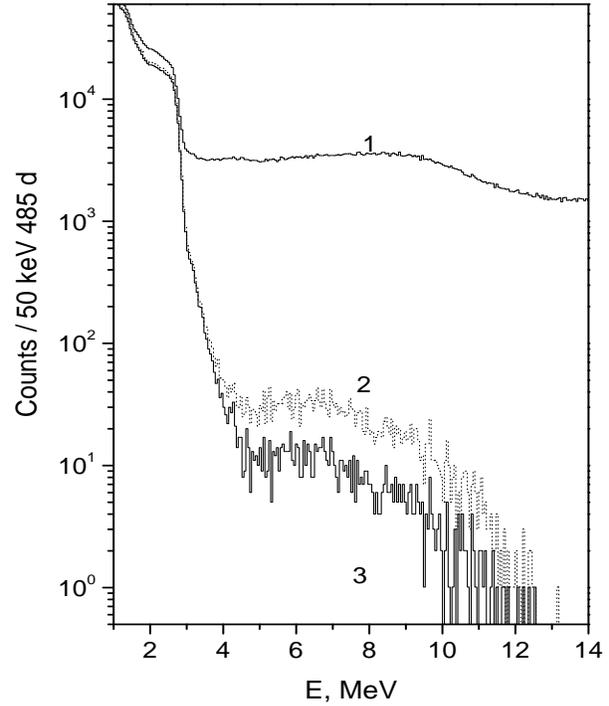}
\caption { Energy spectra of the events and effect of the selection cuts. From top to bottom: (1) raw spectrum; (2) with 2 ms
muon veto cut; (3) with events within 0.7 s of a muon crossing the SSS removed;} \label{Spectr_all}
\end{figure}
The second spectrum is obtained by vetoing all events within 2 ms after muon. The events were selected with the additional requirement that
the mean time of the hits belonging to the cluster with respect to the first hit of the cluster must be $\leq$~100~ns and the time
corresponding to the maximum density of hits must be $\leq$~30 ns. This cut rejects residual muons that were not tagged by the outer water
Cherenkov detector and that interacted in the PC buffer regions. To reduce the background due to the short-lived isotopes ($^9$Li, 178 ms;
$^8$He, 119 ms) induced by muons, an additional 0.7 s veto is applied after each muon crossing the SSS (line 3, Fig.\ref{Spectr_all}). This
cut induces 3.5\% dead time that reduces the live-time to 467.8 days. No events with energy higher than 12.5~MeV passed this cut. This fact
will be used to set limits on the PEP forbidden transitions with $\gamma$- and $\beta^{\pm}$- emissions which have large Q-values (see
Table \ref{Qvalues}).

For PEP forbidden transitions with nucleons emission we analyzed the data in the range (0.5$\div$8.0) MeV. In this energy region
it is necessary to apply a fiducial volume cut in addition to the cuts described above, in order to reject external background.
Fig.{\ref{FV_spectrum} shows the effect of selecting only the innermost 100 tons of scintillator by applying a cut  R = 3.02 m
(line 1).

The spectrum below 3 MeV is significantly suppressed by the fiducial cut, by a factor $\approx$10$^2$. The shape of the
background in the range of (1$\div$2) MeV is determined by cosmogenic $^{11}$C $\beta^+$-decays.  In the next stage of data
selection we removed couples of correlated events falling in a time window of 2 ms (line 2, Fig.\ref{FV_spectrum}). This cut
mainly reject $^{214}$Bi-$^{214}$Po coincidences from the $^{238}$U chain. Finally, a pulse shape-discrimination analysis based
on the Gatti optimal filter \cite{Gat62} is performed to select nucleons. The line (3) shows the events corresponding the
positive Gatti variable (see \cite{Ali08} for more details).

\section{Results}
%%%%%%%%%%%%%%%%%%%%%%%%%%%%%%%%%%%%%%%%%%%%%%%%%%%%%%%%%%%%%%%%%%%%%%%%%%%%%%%%%%%%%%%%%%%%%%%
\subsection{Limits on non-Paulian transitions with emission of $ \gamma$:  ${^{12}C}\rightarrow{^{12}\widetilde{C}}+\gamma$ }          %%%%% GAMMA
%%----------------------------------------------------------------------------------------------
The limit on the probability of the forbidden transitions $^{12}$C$\rightarrow^{12}\widetilde{C}+\gamma$ violating the PEP is based on the
experimental fact that no events above 12.5 MeV survive the selection cuts.

\begin{figure}
\includegraphics[bb = 30 30 500 760, width=8cm,height=11cm]{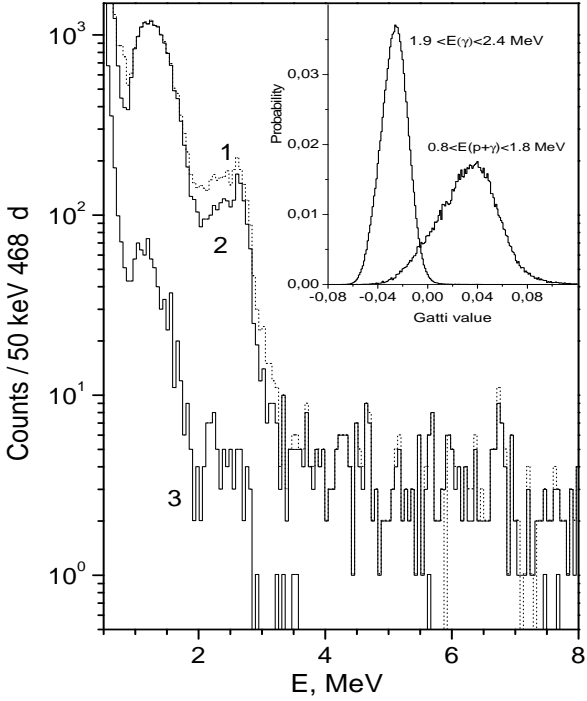}
\caption {The energy spectra of events registered inside the FV (R$\leq$ 3.02 m). (1) spectrum obtained with 2 ms and 0.7 s muon
veto cut; (2) pairs of correlated events (with time interval $\Delta t\leq$ 2 ms between signals) are removed; (3) spectrum of the
events with positive Gatti variable.
 In the inset the values of Gatti variable obtained with $^{241}$Am$^9$Be source for protons and 2.2 $\gamma$ are shown.} \label{FV_spectrum}
\end{figure}

 The lower limits on lifetime for PEP violating transitions of nucleons from $P$-shell to the occupied $1S_{1/2}$-shell were obtained using the formula:
\begin{equation}
 \tau \geq \varepsilon(\Delta E)\frac{N_{N}N_{n}}{S_{lim}}T,
 \label{TLimit}
\end{equation}
where $ \varepsilon(\Delta E) $ is the detection efficiency of an event in the energy interval $ \Delta E $, $ N_{N}$ is the
number of nuclei under consideration, $N_{n}$ is the number of nucleons ($n$ and/or $p$) in the nuclei for which the non-Paulian
transitions are possible, $ T $ is the total time of measurements, and $ S_{lim} $ is the upper limit on the number of candidate
events registered in the $ \Delta E $ energy interval and corresponding to the chosen confidence level.

As shown in Table~\ref{Qvalues}, the most probable energy of $\gamma$-quanta emitted in the nucleon transition from the shell
$1P_{3/2}$ to the shell $1S_{1/2}$ is $\simeq$~17.8~MeV. Taking into account the error of Q-values, the energy of $\gamma$-quanta is inside
the energy interval (16.4$\div$19.4) MeV with 90\% probability. The efficiency of $\gamma$ detection is found for the conservative value
$E_{\gamma}$=16.4 MeV. The response function of the Borexino to the $\gamma$'s of this energy was found by MC simulations based on
GEANT4 code. The uniformly distributed $\gamma$'s were simulated inside the inner vessel (PC + PPO) and in the 1~m- thick layer of buffer
(PC+DMP) surrounding the inner vessel.  The response function is shown in Fig.\ref{response}, the obtained efficiency of 16.4~MeV $ \gamma
$ detection is $ \varepsilon _{\Delta E}=0.50$.

The number of $ ^{12}\rm{C}$ target nuclei in $ 533 $ tons of PC is $ N_{N}=2.37\cdot 10^{31} $ (taking into account the isotopic abundance
of $^{12}$C). The number of nucleons on the $ P $-shell is $ N_{n}=8 $, the total data taking time is $T=1.282$~y, and the upper limit on
the number of candidate events is $S_{lim}=2.44$ with 90\%~C.L. in accordance with the Feldman-Cousins procedure \cite{Fel98}. The limit
obtained using the cited numbers is:
\begin{equation}\label{limgamma}
\tau_{\gamma} (^{12}C\rightarrow ^{12}\widetilde{C}+\gamma )\geq 5.0\cdot10^{31}\; \rm{y}, \label{GLimit}
\end{equation}
for the 90\% c.l. The result improves by more than 4 orders of magnitude our previous limit, obtained with CTF \cite{Bor04}: $\tau
({^{12}C}\rightarrow {^{12}\widetilde{C}}+\gamma )\geq 2.1\cdot10^{27}$ y. This result is stronger than the one obtained with the NEMO-2
detector $\tau ({^{12}C}\rightarrow {^{12}\widetilde{C}}+\gamma )\geq 4.2\cdot 10^{24} $ y \cite{NEMO}, and is comparable with the
Kamiokande detector for $^{16}$O nuclei: $\tau ({^{16}O}\rightarrow {^{16}\widetilde{O}}+\gamma )\geq 1.0\cdot 10^{32}$ y for $\gamma$'s
with energies 19-50 MeV  \cite{KAMIOKANDE}. The limit on the total lifetime of nucleons can be found from the limits on the $\tau_{\gamma}$
as $\tau = \tau_{\gamma}\rm{Br}(\gamma)$, where $\rm{Br}(\gamma) = \Gamma_{\gamma}/\Gamma_{tot}$ is the branching fraction of
$\gamma$-decay. For the case of $^{16}\rm{O}$ nucleus the calculated value of $\rm{Br}(\gamma)$ is inside interval $(2.7-10.4)\cdot10^{-5}$
\cite{KAMIOKANDE}. Unlike the Kamiokande, the Borexino can directly detects the non-Paulian transitions with $p$-, $n$- or
$\alpha$-emission.

\begin{figure}
\includegraphics[bb = 30 30 500 760, width=8cm,height=10cm]{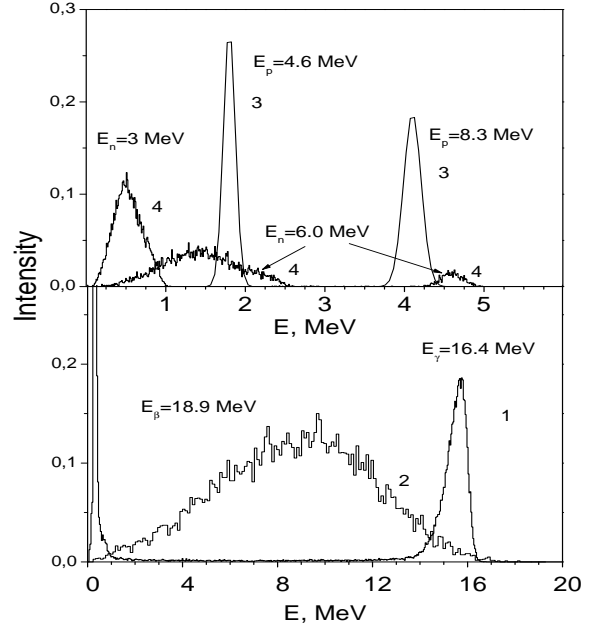}
\caption {The response functions of Borexino: 1) ${^{12}C}\rightarrow{^{12}\widetilde{C}}+\gamma$ (16.4 MeV) decays in IV and
1~m thick layer of buffer; 2) ${^{12}\rm{C}}\rightarrow {^{12}\widetilde{\rm{N}}}+e^{-}+\overline{\nu}$ (18.9 MeV); 3)
${^{12}C}\rightarrow^{11}\widetilde{B}+p$ (4.6 and 8.3 MeV); 4)${^{12}C}\rightarrow{^{11}\widetilde{C}}+n$ (3.0 and 6.0 MeV);}
\label{response}
\end{figure}

\subsection{Limits on non-Paulian transitions in $^{12}$C with proton emission $^{12}$C$\rightarrow ^{11}\widetilde{B}+p$.}          %%%% PROTON
%%---------------------------------------------------------------------------------------------

Using the data of Table~\ref{Qvalues}, one can obtain that the energy released in these transition is within the (5.0-9.0) MeV
interval with a probability of 90\%. Taking into account the recoil energy of ${^{11}\widetilde{B}}$ nucleus, the energy of the
proton is (4.6-8.3) MeV.

The response function of protons were simulated by MC code which takes into account the quenching factor for protons
(Fig.\ref{response}). The empirical Birks' constant \cite{Bir51} was determined from the spectrum of recoil protons measured
with ${^{241}\rm{Am}}{^9\rm{Be}}$-source. It was found that the light yield for a proton with the energy $E_{p}$=4.6(8.3)~MeV
corresponds to an electron energy of $E_{e}$=1.8(4.1)~MeV. It means that the proton peak can be found in the energy interval
1.8-4.1~MeV with 90\% probability. The uncertainty of the peak position is much higher than the energy resolution of the
detector ($ \sigma_E \cong$ 80 keV for $E_e$ = 2~MeV). First, we looked for the proton's peak in the spectrum of single events
obtained with FV cut (line 2, Fig.\ref{FV_spectrum}). The measured spectrum is fitted by polynomial function and gaussian for
proton peak with different positions. Except the region of 2.614 MeV $\gamma$ peak, this procedure gives $S_{lim}=52$ at
90\%~c.l.. The lower limit on the life-time was found from the formula (\ref{TLimit}) taking into account that
$N_N$=4.45$\cdot$10$^{30}$ for 100 tons FV mass:
\begin{equation}\label{limp1}
\tau_p ({^{12}\rm{C}}\rightarrow {^{11}\widetilde{\rm{B}}}+p)\geq 8.9\cdot 10^{29}\; \rm{y}\: (90\%\: c.l.).
\end{equation}

The more stringent limit can be obtained by analyzing the spectrum of signals with positive value of Gatti variable (line 3,
Fig.\ref{FV_spectrum}) that corresponds to the detection of $\alpha$- or proton. The lower limit on lifetime is:
\begin{equation}\label{limp}
\tau_p ({^{12}\rm{C}}\rightarrow {^{11}\widetilde{\rm{B}}}+p)\geq 2.1\cdot 10^{30}\; \rm{y}\: (90\%\: c.l.).
\end{equation}
where the efficiency of Gatti cut $\varepsilon$=0.89 was taken into account. The corresponding upper limits on nuclear instabilities of
$^{12}\rm{C}$ nucleus differ from the limits (\ref{limp1}) and (\ref{limp}) by factor $N_n$=8.

The obtained upper limits on the nuclear instabilities of $^{12}\rm{C}$ are about 4 orders of magnitude stronger than ones obtained with
the 300~kg NaI ELEGANT~V detector $\tau ({^{23}\rm{Na}},{^{127}\rm{I}}\rightarrow
{^{22}\widetilde{\rm{Ne}}},{^{126}\widetilde{\rm{Te}}}+p)\geq 1.7\cdot 10^{25}\; \rm{y}\: (90\%\: c.l.) $ for protons with $E_p \geq$18~MeV
\cite{Ejiri}, and with the 250 kg NaI DAMA/LIBRA detector $ \tau ({^{23}\rm{Na}}, {^{127}\rm{I}}\rightarrow
{^{22}\widetilde{\rm{Ne}}},{^{127}\widetilde{\rm{Te}}}+p)\geq 1.9\cdot 10^{25} \rm{y}\: (90\%\: c.l.)$ for protons with $E_p \geq$10~MeV
\cite{DAMA09}.

The energy of $\alpha$-particles emitted in ${^{12}\rm{C}}\rightarrow ^{8}\widetilde{\rm{Be}}+\alpha$ decay can be found in the
1.0-3.0 MeV interval. Because of the quenching factor, it corresponds to an electron energy range 70-250 keV. The energy 70 keV
is close to the Borexino lower energy threshold and we have not analyzed this reaction with the Borexino data. Our limit on this
mode of transition, which was obtained using the CTF measurements with 20 keV threshold, is $\tau({^{12}\rm{C}}\rightarrow
{^{8}\widetilde{\rm{Be}}}+\alpha) \geq 6.1\cdot 10^{23}$ y (90\% c.l.).

\subsection{Limit on non-Paulian transition in $^{12}$C with neutron emission: ${^{12}\rm{C}}\rightarrow{^{11}\widetilde{\rm{C}}}+n$}     %%%NEUTRON
%%-----------------------------------------------------------------------------------------

Following the calculations of the previous section, one can obtain that the kinetic energy of the initial neutron is in the
(3.2-7.3) MeV interval with 90\% probability. The resulting neutrons are thermalized in hydrogen-rich media of organic
scintillator. The lifetime of neutrons in PC is $\tau \cong 250 ~\mu$s, after which they are captured by protons. The cross
section for the capture on a proton for a thermal neutron is $0.33$~barns. The capture of thermal neutrons $ n+p\rightarrow
d+\gamma $ is followed by $ \gamma $- emission with 2.2~MeV energy. The cross sections are much smaller for capture on $
^{12}\rm{C}$ nuclei($ \sigma _{\gamma}$ = 3.5~mbarns, $E_{\gamma}$ = 4.95 MeV). As a results, the 4.95 MeV peak intensity is
about 1\% from 2.2 MeV peak (Fig.\ref{AmBe}).

The background levels measured in Borexino at 2.2 MeV energy can been used to obtain an upper limit on the number of $ \gamma
$'s with 2.2~MeV energy, and as a result, a limit on the probability of neutron production in the reactions $^{12}$C$\rightarrow
^{11}\widetilde{\rm{C}}+n $. As protons that were scattered during the thermalization can be registered by the detector the
sequential events were not cut out in the data selection (see Fig.\ref{FV_spectrum}, line 1). The response function of the
Borexino to the $\gamma$'s of 2.2~MeV energy was precisely measured with ${^{241}\rm{Am}}{^9\rm{Be}}$ neutron source. The
position and width of the peak is well known, the fitting procedure gives $S_{lim}$=57. Using equation (\ref{TLimit}) one can
obtain the limit on the probability on neutron emission: $\tau_n ({^{12}\rm{C}}\rightarrow {^{11}\widetilde{\rm{C}}}+n)\geq 8.1
\cdot 10^{29} \rm{y}$ (90\% c.l.)

The more stringent limits can be obtained by selecting two consequential  events  inside the full PC volume - the first signal
is from the recoil protons and the second one is 2.2 MeV $\gamma$ from the neutron capture. Candidate events were searched among
all the correlated events occurring within 1.25~ms (5$\tau$) one after another, excluding coincidence times smaller than 20
$\mu$s. The energy of the prompt event was set to be E$\geq$0.5 MeV. The lower threshold is defined by the minimal neutron
energy 3.2 MeV (visible energy of 0.6~MeV) taking into the rate of random coincidences. The response functions for neutrons with
energy 3.0 and 6.0 MeV are shown in Fig.\ref{response}. The energy of the second event was required to be
1.0~MeV$\leq$E$\leq$2.4~MeV for detecting the 2.2~MeV $\gamma$'s with high efficiency. Additionally, the restored positions of
the events have to be within 2 m distance due to the high energy of initial neutron. In such a way, 52 events were selected.
Then for different neutron energies $E_n$ inside (3.2-7.3) MeV interval, the corresponding energy regions for recoil proton
signals were calculated (see lines 4, Fig.\ref{response}). If $E_n$ exceeds the energy of first exited state of $^{12}\rm{C}$
then the high energy part connected with detection of 4.44 MeV $\gamma$'s appears in the spectrum of the prompt events. The
maximal value of the correlated events $N$=26 was found for the ranges (0.6--2.3) MeV and (4.3--5.0) MeV that corresponds 6 MeV
neutrons. Taking into account the probability to find 6.0 MeV neutrons signal in these range ($\varepsilon$=0.9), efficiency of
registering 2.2 MeV $\gamma$'s ($\varepsilon$=0.96), full number of $^{12}$C in the inner vessel $N_N$=1.24$\cdot$10$^{31}$ and
$S_{lim}$=33 for 90\% c.l., the limit is:
\begin{equation}\label{limn}
\tau_n({^{12}\rm{C}}\rightarrow {^{11}\widetilde{\rm{C}}}+n) \geq 3.4\cdot 10^{30} \rm{y} (90\%~c.l.).
\end{equation}

This result is 8 orders of magnitude stronger than the one obtained through searching for spontaneous neutron emission from lead: $
\tau(\rm{Pb}\rightarrow \widetilde{\rm{Pb}}+n)\geq 2.1\cdot 10^{22}\; y\: (68\%\: c.l.) $ \cite{Kishimoto}.

\subsection{Limits on non-Paulian $ \beta ^{\pm}$-transitions: {$ {^{12}\rm{C}}\rightarrow {^{12}\widetilde{\rm{N}}}+e^{-}+\overline{\nu}$} and
$ {^{12}\rm{C}}\rightarrow ^{12}\widetilde{\rm{B}}+e^{+}+\nu $}  %%%%%BETA-
%%------------------------------------------------------------------------------------------------

The energy released in the reaction $^{12}$C$\rightarrow ^{12}\widetilde{N}+e^{-}+\overline{\nu} $ is in (16.4 - 21.4) MeV
interval. The shape of the $ \beta ^{-} $ spectrum with the most probable end-point energy 18.9~MeV is shown in
Fig.\ref{response}. The spectrum was determined by MC method. The limit on the probability of non-Paulian $\beta^-$-transition
was based again on the fact of observing no events with $E_e\geq$12.5~MeV not accompanied by a muon veto signal. The obtained
efficiency of detection of electrons with energies $E_{e}>12.5 $~MeV is $ \varepsilon =0.12 $. The limit on the lifetime of
neutrons ($N_n$=4) in $^{12}$C with respect to the transitions violating the PEP is
\begin{equation}\label{limbeta-}
\tau_{\beta^-} (^{12}\rm{C}\rightarrow ^{12}\widetilde{\rm{N}}+e^{-}+\overline{\nu} )\geq 3.1\cdot 10^{30}\: \rm{y}\:
(90\%~c.l.).
\end{equation}
This result is 6 orders of magnitude stronger than the one obtained by NEMO-2, $\tau({^{12}\rm{C}}\rightarrow
{^{12}\widetilde{\rm{N}}}+e^{-}+\overline{\nu} )\geq 3.1\cdot 10^{24}\: \rm{y}\: (90\%\: c.l.) $ \cite{NEMO}.

The data available from the LSD detector \cite{LSD} situated in the tunnel under Mont Blanc allows obtaining a qualitative limit
for this decay mode. In \cite{Kek90}, it is claimed that only 2~events were observed with energies higher than 12~MeV during
75~days of data taking with the detector loaded with 7.2~tons of scintillator, containing $ 3\times 10^{29} $ $^{12}$C nuclei.
The upper limit that can be obtained using formula (\ref{TLimit}) with these data (with $S_{lim}$=5.91 events for 90\%~c.l. and
detection efficiency $ {\varepsilon (\rm{E}\geq \rm{12~MeV})} =0.23 $ is $\tau (^{12}$C$\rightarrow
^{12}\widetilde{\rm{N}}+e^{-}+\overline{\nu} )\geq  9.5\cdot10^{27}$ y (90\%~c.l.).

The end-point energy of the $ \beta ^{+} $ spectrum is 16.8~MeV, but the spectrum is shifted towards higher energies by $ \simeq
0.85 $~MeV due to the registering of annihilation quanta (Fig.\ref{response}). The efficiency of the $ ^{12}$C$\rightarrow
^{12}\widetilde{\rm{B}}+e^{+}+\nu $ transition detection with energy release $ E>12.5 $~MeV is $ \varepsilon =0.079 $. The lower
limit on the lifetime of the proton in the $^{12}$C nuclei is then
\begin{equation}\label{limbeta+}
\tau_{\beta^+} ({^{12}\rm{C}}\rightarrow {^{12}\widetilde{\rm{B}}}+e^{+}+\nu )\geq 2.1\cdot 10^{30}\: \rm{y}\: (90\%\: c.l.)
\end{equation}
The limits obtained by the NEMO-2 collaboration for this reaction are 6 orders of magnitude weaker: $ \tau
(^{12}\rm{C}\rightarrow ^{12}\widetilde{\rm{B}}+e^{+}+\nu )\geq 2.6\cdot 10^{24}\: y\: (90\%\: c.l.) $ \cite{NEMO}.

The final limits on the nucleon instability are shown in Table~\ref{Lim} in comparison with the the previous results obtained
for the same PEP-violating transitions . The limit \cite{DAMA09} relates to the instability of $^{23}\rm{Na}$ and $^{127}\rm{I}$
nuclei, all other limits are given per nucleon for which the non-Paulian transition is possible.

\begin{table}[!htbp]
\begin{center}
\caption{Mean lifetime limits for non-Paulian transitions of nucleons in the Borexino.} \label{Lim}
\begin{tabular}{|l|c|c|c|}
  \hline
    Channel & $\tau_{lim}$ (y) & Previous\\
             &         90\% c.l.         & limits \\ \hline
   $^{12}\rm{C}\rightarrow^{12}\widetilde{\rm{C}}+\gamma$ & 5.0$\cdot10^{31}$ & 4.2$\cdot10^{24}$($^{12}\rm{C}$)\cite{NEMO}  \\
                                                           &                  & 1.0$\cdot10^{32}$($^{16}\rm{O}$)\cite{KAMIOKANDE}  \\
   $^{12}\rm{C}\rightarrow^{11}\widetilde{\rm{B}}+ p$     &  8.9$\cdot10^{29}$ & 1.9$\cdot10^{25}$($^{23}\rm{Na},^{127}\rm{I}$)\cite{DAMA09}  \\
   $^{12}\rm{C}\rightarrow^{11}\widetilde{\rm{C}}+ n$    & 3.4$\cdot10^{30}$ & 2.1$\cdot10^{22}$($^{nat}\rm{Pb}$)\cite{Kishimoto}  \\
   $^{12}\rm{C}\rightarrow^{12}\widetilde{\rm{N}}+ e^{-}+\overline{\nu_{e}}$ & 3.1$\cdot10^{30}$ & 9.5$\cdot10^{27}$($^{12}\rm{C}$) \cite{Kek90,LSD}\\
   $^{12}\rm{C}\rightarrow^{12}\widetilde{\rm{B}}+ e^{+}+\nu_{e}$ & 2.1$\cdot10^{30}$ & 2.6$\cdot10^{24}$($^{12}\rm{C}$)\cite{NEMO} \\
   \hline
 \end{tabular}
\end{center}
\end{table}

\subsection{Limits on the relative strength of non-Paulian transitions}
%%%%%%%%%%%%%%%%%%%%%%%%%%%%%%%%%%%%%%%%%%%%%%%%%%%%%%%%%%%%%%%%%%%%%%%%%%%%%%%%%%%%%%%%%%%%%%%%%%%%%%%%%%%%%%%%%%%%%%%%%

The PEP forbidden transitions with emission of $\gamma$-, $n$- or p- and $\nu$,e-pair can be induced by electromagnetic, strong
and weak interactions, correspondingly. The obtained upper limits on lifetime for different processes can be converted to limits
on the relative strength of non-Paulian transitions to the normal one: $\delta^2$= $\widetilde{\lambda}/\lambda$, where
${\lambda} = 1/\tau$ is unit time probability (rate) of forbidden ($\widetilde{\lambda}$) and normal (${\lambda}$) transitions.
The ratio $\delta^2$=$(g_{PV}/g_{NT})^2$ is a measure of the violation of the PEP and represents the mixing probability of
non-fermion statistics allowing the transitions to the occupied states. In particular, in quon model of PEP-violation
\cite{Gre90,Moh90} the parameter $\delta^2 = \beta^2/2$ corresponds to the probability of admixed symmetric component of the
particle. In this way one can compare the experimental limits on the lifetime obtained for different nuclei and atoms.

The decay width of the nuclear electric dipole 16.4 MeV E1 $\gamma$ -transition from $P$- to $S$-shell given by the Weisskopf estimate is
$\rm{\Gamma_{\gamma}}\approx$ 1.5 keV, then the rate of normal E1 transition is $\lambda = \rm{\Gamma_{\gamma}}/ \hbar$=
2.3$\cdot$10$^{18}$ s$^{-1}$. With the obtained upper limit on $\tau_{\gamma}$ (\ref{GLimit}), the ratio
$\delta^2_{\gamma}=\widetilde{\lambda}(^{12}\rm{C})/\lambda(^{12}\rm{C}) $ is less than 2.2$\cdot 10^{-57}$ (90\%c.l.) This limit is close
to result of Kamiokande detector for $^{16}$O nuclei - $\delta^2_{\gamma}=2.3\cdot10^{-57}$ \cite{KAMIOKANDE}.

Although the E1-transition is the fastest among $\gamma$-transitions, the width of hadron emissions is 3-4 orders larger than that of
$\gamma$-transitions. The widths of single S-hole states in $^{12}$C measured for $(p,2p)$- and $(p,pn)$-reaction  are $\rm{\Gamma_{n,p}}
\cong$ 12 MeV \cite{PNPI}. As a result, the detection of protons or neutrons gives a more stringent limit on the relative strength of PEP
forbidden transitions than the detection of $\gamma$'s if one can set a similar limit on the lifetime for both decays. Using the lower
limits on $\tau_p$ (\ref{limp}) and $\tau_n$ (\ref{limn}) one can obtain the limits $\delta^2_{p}=\widetilde{\lambda}/\lambda \leq $
1.6$\cdot 10^{-59}$ and $\delta^2_{n}\leq $ 4.1$\cdot 10^{-60}$ at 90\% c.l.. This result is more then 4 orders of magnitude stronger than
the one obtained by DAMA collaboration \cite{DAMA09}.

The non-Paulian $\beta ^{\pm}$ -transitions are first-order forbidden $P\rightarrow S$ transitions. The $log(ft_{1/2})$ values
for such first forbidden transitions is 7.5$\pm$1.5.  The conservative value $log(ft_{1/2})$=9 corresponds to life-time
$\tau\approx$ 480 sec for Q=18.9 MeV in the case of $\beta^-$-decay (level width is $\rm{\Gamma}_{\beta^-}\approx$
1.4$\cdot10^{-18}$ eV) and $\tau\approx$ 1050 sec (Q=17.8 MeV, $\beta^+$). As result, the restrictions on the relative strength
of non-Paulian $\beta^{\pm}$-decays are significantly lower: $\delta^2_{\beta^-}\leq $ 2.1$\cdot 10^{-35}$ and
$\delta^2_{\beta^+}\leq $ 6.4$\cdot 10^{-35}$ (90\%~c.l.). The previous  limit $\delta^2_{\beta^-}\leq $ 6.5$\cdot 10^{-34}$
obtained in \cite{Kek90} with LSD data \cite{LSD} is in 30 times weaker. It should be noted, that although the limit for
$\beta^{\pm}$-transitions are significantly lower than the nucleon ones, there is a significant difference between these
processes. As mentioned above in $\beta^{\pm}$ decays new particle ($p$ or $n$) arises in non-Paulian state, thus
Amado-Primakoff arguments may not be valid \cite{Ama80,Kek90}. The limit on $\delta^2_{\beta^{\pm}}$ can be compared with the
limit obtained by the VIP experiment - $\beta^2/2 \leq$4.5$\cdot 10^{-28}$ \cite{Bar06}.

The upper limits obtained on the relative strengths of non-Paulian transitions are shown in Table~\ref{RelStr}. For transitions
with ($n,p$)- and $\beta^{\pm}$-emission the stronger limit is included.

\begin{table}[!htbp]
\begin{center}
\caption{Upper limits on the relative strength, $\delta^2=\widetilde{\lambda}/\lambda$ (at 90\% C.L.), for non-Paulian
transitions in the Borexino.} \label{RelStr}
\begin{tabular}{|l|c|c|c|c|}
  \hline
    decay  & $\widetilde{\lambda}(^{12}\rm{C})$, & $\lambda(^{12}\rm{C})$ & $\delta^2=\widetilde{\lambda}/\lambda$  & Previous\\
             &         (s$^{-1}$)     & (s$^{-1}$)           &                                                      & limits \\ \hline
   $\gamma$  & 5.0$\cdot$10$^{-39}$   & 2.3$\cdot$10$^{18}$ & $$2.2$\cdot$10$^{-57}$     & 2.3$\cdot10^{-57}$\cite{KAMIOKANDE}  \\
     N(n,p)  & 7.4$\cdot$10$^{-38}$   & 1.8$\cdot$10$^{22}$ & $$4.1$\cdot$10$^{-60}$     &  3.5$\cdot10^{-55}$\cite{DAMA09} \\
  ($ e,\nu$) & 4.1$\cdot$10$^{-38}$   & 2.0$\cdot$10$^{-3}$ & $$2.1$\cdot$10$^{-35}$     & 6.5$\cdot10^{-34}$ \cite{Kek90,LSD}\\
   \hline
 \end{tabular}
\end{center}
\end{table}

\section{Conclusions}
%%%%%%%%%%%%%%%%%%%%%%%%%%%%%%%%%%%%%%%%%%%%%%%%%%%%%%%%%%%%%%%%%%%%%%%%%%%%%%%%%%%%%%%%%%%%%%%

Using the unique features of the Borexino detector -- extremely low background, large scintillator mass of 278 tons, low energy
threshold and a carefully designed muon-veto system -- the following new limits on non-Paulian transitions of nucleons from the
$1P_{3/2}$-shell to the $1S_{1/2}$-shell in $^{12}$C with the emission of $\gamma, n, p$ and $\beta^{\pm}$ particles have been
obtained:
\begin{center}

\noindent $\tau(^{12}$C$\rightarrow^{12}\widetilde{\rm{C}}+\gamma) \geq 5.0\cdot10^{31}$~y,

\noindent $\tau(^{12}$C$\rightarrow^{11}\widetilde{\rm{B}}+ p) \geq 8.9\cdot10^{29}$~y,

\noindent
 $\tau(^{12}$C$\rightarrow^{11}\widetilde{\rm{C}}+ n) \geq 3.4 \cdot10^{30}$~y,

\noindent $\tau(^{12}$C$\rightarrow^{12}\widetilde{\rm{N}}+ e^- + \nu) \geq 3.1\cdot10^{30}$~y
%\noindent
\end{center}
and \begin{center} \noindent $\tau(^{12}\rm{C}\rightarrow^{12}\widetilde{\rm{B}}+ e^+ + \overline{\nu}) \geq 2.1\cdot10^{30}$ y,
\end{center}
all with 90\% C.L.

Comparing these values with the data of Table~\ref{Lim}, one can see that these limits for non-Paulian transitions in $^{12}$C
with $\gamma$-,~$p$-,~$n$-, and $\beta^{\pm}$- emissions are the best to date. The obtained lifetime limits allow to introduce
the new upper limits on the relative strengths of the non-Paulian transitions to the normal ones: $\delta^2_{\gamma}\leq
2.2\cdot 10^{-57}$, $\delta^2_{N}\leq 4.1\cdot 10^{-60}$ and $\delta^2_{\beta}\leq 2.1\cdot 10^{-35}$, all at 90\% c.l.

\section{Acknowledgements}
%%%%%%%%%%%%%%%%%%%%%%%%%%%%%%%%%%%%%%%%%%%%%%%%%%%%%%%%%%%%%%%%%%%%%%%%%%%%%%%%%%%%%%%
 The Borexino program was made possible by funding from INFN (Italy), NSF (U.S.), BMBF, DFG and MPG (Germany), Rosnauka (Russia), and MNiSW (Poland).
 We acknowledge the generous support of the Laboratori Nazionali del Gran Sasso (LNGS). A. Derbin acknowledges the support of Fondazione Cariplo.

\end{document}